\documentstyle{article}

\def\oe{\omega^\varepsilon} 
\def\bR{{\bf R}}    \def\bZ{{\bf Z}}
\let\a=\alpha  \let\g=\gamma \let\de=\delta
\let\e=\varepsilon   
  \let\la=\lambda 
  \let\r=\rho

\let\La=\Lambda  \let\De=\Delta

\def\lb{\left(} \def\rb{\right)}
 
\def\0#1#2{\frac{#1}{#2}}
\def\s0#1#2{\mbox{\small{$\frac{#1}{#2}$}}}
\def\5{\bar }  \def\p{\partial } \def\7{\hat } \def\4{\tilde }

\let\da=\downarrow

\let\ol=\overline

\def\bea{\begin{eqnarray}} \def\eea{\end{eqnarray}}
\def\beann{\begin{eqnarray*}} \def\eeann{\end{eqnarray*}}
\def\beq{\begin{equation}} \def\eeq{\end{equation}}
\def\ba{\begin{array}} \def\ea{\end{array}}

\newcommand{\mysection}[1]{\section{#1}\setcounter{equation}{0}
\setcounter{theorem}{0}\setcounter{lemma}{0}\setcounter{corollary}{0}}

\def\ben{\begin{enumerate}}
\def\een{\end{enumerate}}

\newtheorem{prop}{Proposition}
\newtheorem{definition}{Definition}

\begin{document}


\thispagestyle{empty}
\begin{flushright}
      UWThPh-1997-31
\end{flushright}
\vfill
\begin{center}{\Large
\bf Hydrodynamics for\\[1ex]
    Quasi-Free
    Quantum Systems\\[1ex]}
\end{center}
\vfill

\begin{center}
{\Large
Christian Maes\footnotemark }\\*[.5em]
Instituut voor Theoretische Fysica,\\
Katholieke Universiteit Leuven,\\
Celestijnenlaan 200 D, B--3001 Leuven, Belgium

\vfill
and
\vfill
{\Large Wolfgang Spitzer\footnotemark$^,$\footnotemark}\\*[.5em]
Institut f\"ur Theoretische Physik,\\
Universit\"at Wien,\\
Boltzmanngasse 9, A--1090 Wien, Austria

\end{center}

\vfill
\begin{abstract} We consider quasi-free quantum systems and
we derive the Euler equation using the so-called hydrodynamic
limit. We use Wigner's well-known distribution function and discuss an
extension to band distribution functions for particles in a periodic potential.
We investigate the Bosonic system of hard rods and calculate
fluctuations of the density.

\end{abstract}
\vfill
{\bf Keywords:} Euler equation, quantum distribution function,
                hydrodynamic limit
\vfill
\addtocounter{footnote}{-3}
\addtocounter{footnote}{1}
\footnotetext{Onderzoeksleider F.W.O. Flanders
\\ E-mail address: Christian.Maes@fys.kuleuven.ac.be}
\addtocounter{footnote}{1}
\footnotetext{Supported by the project P10517-NAW of the \"Osterreichische
Forschungsfond, FWF}
\addtocounter{footnote}{1}
\footnotetext{
Present address: Department of Mathematics, University of Copenhagen, 
Universitetsparken 5, DK-2100 Copenhagen, Denmark.
E-mail: spitzer@math.ku.dk}

\newpage
\setcounter{page}{1}
\setcounter{footnote}{0}

\mysection{Introduction}

The Euler equations appear in hydrodynamics as scaling limits for the
dynamics of the conserved quantities in a fluid.  So far the only rigorous
example of a microscopic derivation starting from Newton's laws is that of
the one-dimensional hard rod fluid, \cite{BDS}, \cite{Dob}, \cite{S1},
\cite{BS}. In this degenerate situation the number of
particles with a given momentum is locally conserved.  Therefore the
hydrodynamic field $n_v(x,t)$ is just counting the number of particles
at  time $t$ with space coordinate $x$ and with momenta equal to $v$ (we
have put the masses equal to one). When the rod length tends to zero we
deal with a one-dimensional ideal gas and, after the appropriate (Eulerian)
scaling, the associated density field converges to the solution of
\beq\label{ideal}
     \p_t f(x,v;t) + v \p_x f(x,v;t)=0
\eeq
(for certain initial conditions).  One can then go further and study the
fluctuations around this macroscopic equation.  They are Gaussian with a
covariance matrix containing information about the transport coefficients
which are finite and non-zero for the hard rod fluid.

In the present paper we investigate what remains of this in the quantum
case.  Then also here we pick out the simplest possible quantum dynamics
and we simply ask how the corresponding Euler equations and the
fluctuations around it can be derived. We see
this as a modest step in the rigorous study of what could be called
quantum hydrodynamics.  Clearly, a mathematical derivation {\it ab
initio} of the macroscopic equations
of so-called quantum liquids is beyond reach. We may however learn
something about the conceptual set-up to start such a
project from a rigorous study in the most simple cases. We refer to
\cite{S1}, \cite{S2}
for an overview and some solutions to some of the questions in the quantum
domain of hydrodynamics.  In all cases the underlying philosophy is
however not fundamentally different from
the classical situation.  Conserved quantities vary on a much larger
time scale than the others and by a law of large numbers the macroscopic
equations appear as closed equations governing the motion of the rescaled
conserved quantities.

Here we study quasi-free systems.  That means that the microscopic dynamics
is in some sense linear (as for classical Gaussian systems) and
the equations are determined by the
one-particle motion.  Therefore we do not have to deal with the most
important  problem
of time-scale separation (the Boltzmann-Gibbs principle) to see how the
rescaling effectively truncates the hierarchy of kinetic equations.
In this situation two papers \cite{S2} and \cite{S3} by Herbert Spohn show
how to get the program started. In the next Section we summarize some of
the ideas.  In Sections 3 and 4 we discuss the derivation of the Euler
equation (similar to (\ref{ideal})) for free systems subject to a slowly
varying external potential.  It is interesting to see that this can be
applied (in Section 5) in the case of hard core bosons on
the one-dimensional lattice.  Section 6 deals with the
non-equilibrium fluctuations.  We compute the covariance of the
rescaled density field. As expected, only at that moment do we start
seeing a difference between the quantum statistics. Finally, in
Section 7, we turn to the case of non-interacting quantum particles
in a periodic potential.

\mysection{ Some Preliminaries}

The first step in doing anything related to quantum hydrodynamics is
probably asking for what quantum systems we can get a good idea of what
are the conserved quantities and what is the structure of the equilibrium
states. The answer is disappointing: compared to the situation in so-called
classical statistical mechanics, there is very little we know about
truly interacting quantum systems.
It is therefore not unreasonable to turn first to quasi-free systems where
already on the microscopic scale the dynamics is governed by a
one-particle Hamiltonian.  A quasi-free dynamics is  a quantum dynamics
generated by a Hamiltonian $H$ which is quadratic in the creation and
annihilation operators $a^*(x), a(x)$ of a Bose or Fermi field.
Formally,
\beq\label{quasi}
H=\int dx dy\, h_\e(x,y) a^*(x)a(y).
\eeq

As usual with a varying number of particles, this should be thought of as
acting on Fock space with single particle space $L^2(\bR^d, dx)$.   The
quadratic form is specified by a one-particle Hamiltonian $h_\e$ which
in the physically more interesting situations has the form
\beq\label{model}
  h_\e = \0 1 2 [\vec{p} + e \vec{A}(\e x)]^2 + U(x) + V(\e x)
\eeq
where $U$ is a periodic potential, $V$ is an external potential and
$\vec{A}$
is the vector potential.
The parameter $\epsilon > 0$ specifies the scale over which the external
potentials are varying and will be our scaling parameter
in what follows. In the next Section we
start with the case where $U=0$ but allowing a bit more general form of
$h_\e(x,y)$ than obtained from (\ref{model}).  It is not too difficult to
see that, when dealing with a quantum dynamics generated by (\ref{quasi}),
the evolution of the correlation functions can be specified in terms of the
single particle evolution generated by $h_\e$. Therefore, the kinetic
equations are closed even on the microscopic level which is an enormous
advantage for doing hydrodynamics.

Notice however that the form
(\ref{quasi}) need not be restricted to the microscopic
domain (where (\ref{quasi})) is rather simplistic); there
are plenty of quasi-free models in mesoscopic physics where (\ref{quasi})
appears as effective action or as describing the evolution of
quasi-particles. E.g. the central question in Bogoliubov's work on
superfluidity was to see how to arrive at something like (\ref{quasi})
(with some very particular properties) starting from  a realistic
interaction between He$^4$-particles, \cite{Bog}, \cite{AVZ}. This is
just a question of equilibrium statistical mechanics.  Once arrived
at (\ref{quasi}) with an energy spectrum $E(k)\sim |k|$ for small $|k|$,
Landau's theory (as e.g. in \cite{LP}) takes over and completes
the dynamical picture of superfluidity.

A quasi-free state is a state on the CAR or CCR algebra
(the algebra's generated by the identity and the $a^*(x), a(x)$ satisfying
the Canonical (Anti)Commutation Relations)
for which all correlation functions (and hence the state itself)
are determined by the two-point functions, see e.g. \cite{BR}.  For gauge
invariant quasi-free states $\omega$ we must only specify the
$\omega(a^*(x) a(y))$, which, in turn, are given in terms of a self-adjoint
operator on the one-particle space.

Because of the dependence on the one-particle evolution in our systems (as
explained above) it is crucial to understand the scaling of the long time
behavior of a quantum particle subject to (\ref{model}).  To keep it
simple, let us for a moment put $A = 0$ in (\ref{model}).   Consider
the position operator $r(t)$ in the Heisenberg picture and define its
rescaling as
\beq\label{scaled}
  r^\e(t) = \e r(\e^{-1} t)
\eeq
Rescaling the momentum operator as $p^\e(t) = p(\e^{-1} t)$ we see that the
commutator $[r^\e,p^\e]$ vanishes as $\e$ goes to zero. Let us therefore
consider the semiclassical equations of motion, cf. \cite{GMMP}, \cite{K},
\cite{MMP}, \cite{S3}. Writing $E_n(k)$ ($n=1,2,\ldots$ and $k\in$ the
Brillouin zone), for the (Bloch-)eigenvalues of the periodic part $h^o
= \0 1 {2} p^2 + U(x)$ (with periodic boundary conditions), for
each band the semiclassical equations ($\hbar = 1$) look as follows:
\bea \label{semi1}
     \p_t r&=&\p_k E_n(k),\\
     \label{semi2}
     \p_t k&=&-\p_r V(r).
\eea
If at time $t=0$ the particle has a probability distribution $f_n(r,k)$,
then, via (\ref{semi1}-\ref{semi2}), the distribution at time $t\geq 0$ is
given by $f_n(r,k;t) = f_n(r(-t),k(-t))$ and solves
\beq \label{semi3}
     \p_t f_n(r,k;t)+\p_k E_n(k)\p_r f_n(r,k;t)=\p_r V(r) \p_k
f_n(r,k;t).
\eeq
We can therefore expect that (\ref{semi3}) gives the correct Euler equation
for the considered quasi-free system but we still need to understand
what to use for distribution function $f_n(r,k;t)$.

The next question is thus to see what is the analogue of classical
quantities like $n_v(x,t)$ or $f(x,v;t)$ appearing in
(\ref{ideal}). Let us therefore briefly recall the notion of Wigner
distribution function (cf.\cite{Wig}). A recent and to this work
very relevant mathematical survey is contained in Section 1 of
\cite{GMMP}. If $\psi(r)$ is the wave function of a quantum system,
then we call ($h=1$):
\beq\label{wigner}
  F(r,k)=\lb\0 1 {\pi}\rb^d \int_{\bR^d} d\eta\,
  e^{2ik\cdot\eta} \bar{\psi}(r+\eta)\psi(r-\eta),
\eeq
the distribution function of the simultaneous values of the
coordinates $r$ and momenta $k$. Even though $F$ is not positive in
general (but it is real valued), this is traditionally justified by the
following properties: \bea
\label{aaa} |\psi(r)|^2&=&\int dk\, F(r,k),\\
\label{bbb} |\tilde{\psi}(k)|^2&=& \int dr\, F(r,k).
\eea
By $\tilde{}$ we denote Fourier transformation.

We wish to do exactly the same thing for our quasi-free systems.  We
will come back to the case of a periodic potential in Section 7 but for the
moment we now also put $U=0$.  Now we have, for every $\e>0$ small, a state
$\oe$ and we
must rescale space and time by $\e^{-1}$. Let $\a_t$ denote the microscopic
time evolution to be generated by a quasi-free Hamiltonian and write
the time-evolved state as $\oe\circ\a_{\e^{-1}t}=\oe_{\e^{-1}t}$.
Then we define (cf.\cite{S2}),

\beq\label{def1}
  f^\e(r,k;t)=
          \lb \0 1 \pi\rb^d\lb\sum_\eta\rb
           \int d\eta\, e^{2i\eta\cdot k}\oe_{\e^{-1}t}
           [a^*(\e^{-1}r+\eta)a(\e^{-1}r-\eta)]
\eeq
as the macroscopic (scale $\e^{-1}$) one-particle distribution
function of the system.  Clearly, clustering conditions on the
state $\oe$ will be required to allow the convergence of the
integral (sum) over $\eta$. From now on the assumption stands that
this clustering holds allowing a well defined (\ref{def1}). Even
though we write an integral in (\ref{def1}) we prefer to avoid here
technicalities related to working in the continuum. We therefore
mostly think of lattice systems (in which case we really have a
discrete sum, $\e^{-1}r$ should be replaced by its integral part
and the wave vector $k$ is in the first Brillouin zone).
Expressions like $a^*(x)a(y)$ should be understood in the
distributional sense.  Also later we will use the integral sign as
a common symbol even when considering discrete systems. The
analogous properties to (\ref{aaa})-(\ref{bbb}) apply. As an
example, note that if we are interested say in the density and if
initially (at $t=0$) we have a product state with
$\oe[a^*(\e^{-1}r+\eta)a(\e^{-1}r-\eta)] = \delta(\eta) \rho(r)$,
then $f^\epsilon(r,k;0)=\0 1 {2\pi}\rho(r)$ is the initial particle
density.

The main question is now to investigate the limiting behavior (as $\e$
goes to zero) of (\ref{def1}). That is, to derive what corresponds to the
Euler equation (\ref{ideal}).  Afterwards we look for the fluctuations
around this limiting behavior.

It should be clear by now that we do not identify the problem of deriving
`quantum' Euler equations or of studying quantum hydrodynamics with that of
`quantizing' the classical hydrodynamic equations.  The macroscopic
equations remain just `ordinary' partial differential equations for
conserved quantities (and not for operators).  Still, the quantum nature of
the underlying
system may in principle have a non-trivial effect on these equations (as
for quantum liquids, superfluidity etc.); if not on the macroscopic
equations themselves then on the fluctuations (with corresponding transport
properties) around it.

Scaling limits for Wigner functions (measures) are studied in the
mathematics literature (see for instance the recent work by
\cite{GMMP}). The starting point there is the Schr\"odinger equation

\beq \label{SE}
 \partial_{t} \psi^{\e} (\e^{-1}x,\e^{-1}t) +i
 h_{\e}\psi^{\e}(\e^{-1}x,\e^{-1}t) =0
\eeq
with $\psi^{\e}(x,t=0)=\psi^{\e}(x)$\footnotemark
\footnotetext{Regarding notation, we identify the $u^{\e}(\e x,\e t)$ of
\cite{GMMP} with our $\psi^{\e}(x,t)$}. $h_{\e}$ is the same type
of Hamilton operator as in our quasi-free systems. In fact, we
shall use their result to show the existence of $f(r,k;t)$, the
limit of (\ref{def1}) as $\e$ goes to zero. Moreover, to a large
extent hydrodynamics of quasi-free quantum systems reduces to the
homogenization limits considered in \cite{GMMP}.

\mysection{Euler Equation for Quasi-Free Systems}

We consider a system of Bosons or of Fermions with formal Hamiltonian
\[H=\int dx dy\, h_\e(x,y) a^*(x)a(y).
\]
As usual, the field of creation and annihilation operators is
denoted by $a^*(x),a(x)$.   The kernel $h_{\e}(x,y)$ corresponds to
a one-particle Hamiltonian.   We always assume here that $h_\e(x,y)
= h(\e(x+y)/2,y-x)$ with an appropriate decay condition, e.g.
$\sup_x \int dy |h_\e(x,y)| < \infty$ for Fermions, to generate an
infinite volume quasi-free time evolution further denoted by
$\a_t$.  Or put differently, the matrix elements $h_{\e}(x-\0 y 2,
x+ \0 y 2) = h(\e x, y)$ may be thought of as hopping rates (from
place $x-\0 y 2$ to $x+\0 y 2$ by a distance $y$) varying slowly
with $x$. The time $t$ is real.  Equivalently, we may write the
Hamiltonian $H$ via the Fourier transform $\tilde{a}(p)=\lb \0 1
{2\pi}\rb^{d/2}\int du\,e^{ipu} a(u)$,
\[H = \lb \0 1 {2\pi}\rb^d \int dk \int dp \int du \,e^{i(k-p)u} E(\e u, (k+p)/2)
      \tilde{a}^*(k)\tilde{a}(p),
\]
where the energy spectrum, $E$, is defined by: $E(r,k)=\int dv\, e^{-i k v} h(r,v)$.
Clearly, in the limit of
infinite scale separation, $\e$ to zero, the microscopic time evolution is
translation invariant with (true) energy spectrum $E(k)=E(0,k)$.  Yet, as
we will see shortly, the (macroscopic) Euler equation remembers the
dependence of $E(r,k)$ on $r$. Finally, the simplest example corresponds
to a free system in a slowly varying chemical potential, $E(r,k) = E(k) +
V(r)$, which corresponds to
\[H=\int dk  E(k) \tilde{a}^*(k) \tilde{a}(k) + \int du V(\e u) a^*(u)
a(u). \]


In order to find conditions under which the existence of the
limiting $f$ (see (\ref{def1})) is guaranteed we recall the work by
\cite{GMMP}. Suppose the 2-point (which is the only relevant
information on the state) function of $\oe$ is of the (general)
form

\beq \label{decomp}\oe [a^{*}(x) a(y)] = \int d\mu(\la)\,
     \bar{\psi}^{\e}_{\lambda}(x) {\psi}^{\e}_{\lambda}(y)
\eeq
where ${\psi}^{\e}_{\lambda}$ satisfies the Schr\"odinger equation
(\ref{SE}). $d\mu(\la)$ is an absolutely integrable signed measure.
On $h_{\e}$ and ${\psi}^{\e}_{\lambda}$ we assume the same
conditions as in \cite{GMMP}.   This decomposition includes the
closed convex hull of pure states (indexed by $\lambda$).  Note,
that

\beq f^{\e}(r,k;t) = \int d\mu(\la)\,
     w^{\e}[{\psi}^{\e}_{\la}] (x,k;t)
\eeq
where $w^{\e}[{\psi}^{\e}_{\la}] (x,k;t)$ is the Wigner function
used by \cite{GMMP}. This relation provides an alternative way to
derive the Euler equation by quoting their result on
$w^{\e}[{\psi}^{\e}_{\la}] (x,k;t)$.

We write $f(r,k;t)$ for any limit  $f(r,k;t) =
\lim_{\e \downarrow 0} f^\e(r,k;t),  \p_s f(r,k;t) =
\lim_{\e\da0}\p_s f^\e(r,k;t), s=t,r,k$ and we wish to see
what equation is satisfied by such a limiting function.

\begin{prop}\label{prop1}
Under the above hypotheses,
\beq \label{euler}
     \p_t f(r,k;t)+\p_k E(r,k)\p_r f(r,k;t)=\p_r E(r,k)\p_k f(r,k;t).
\eeq
\end{prop}

\noindent{\bf Proof:} There are different ways to verify (\ref{euler}).
Below we present an explicit and detailed computation.

\beann
\p_t f^\e(r,k;t)
&=&\lb \0 1 {2\pi}\rb^{d}\int d\eta \, e^{i\eta k}\p_t
   \Big[\oe_{\e^{-1}t}[a^*(\e^{-1}r+\0 \eta 2)a(\e^{-1}r-\0 \eta 2)]\Big]
\\
&=&\0 {-i} \e \lb \0 1 {2\pi}\rb^{3d}\int d\eta \, e^{i\eta k}\int dp dp'
   e^{ip\e^{-1}r-ip'\e^{-1}r+i \0 \eta 2 (p+p')}\\
&&\times\int dx dy dz \,
  e^{-ipx+ip'y}\Big[{h_\e(x,z)}\oe_{\e^{-1}t}[a^*(z)a(y)]-
  \ol{h_\e(y,z)}\oe_{\e^{-1}t}[a^*(x)a(z)]\Big]
\\
&=&\0 {-i} \e \lb \0 1 {2\pi}\rb^{3d}
   \int d\eta dp dp' \,e^{i\eta(k+\0 {p+p'} 2) +i(p-p')\e^{-1}r}
\\
&&\times \left[\int dy dn dm \,e^{ip'y-ipm-ipn}\oe_{\e^{-1}t}[a^*(m)a(y)]
  {h_\e(n+m,m)} \right.
\\
&&-\left.\int dx dn dm \,e^{-ipx+ip'm+ip'n}\oe_{\e^{-1}t}[a^*(x)a(m)]
  \ol{h_\e(n+m,m)}
  \right]
\\
&=&2^d\0 i \e \lb \0 1 {2\pi}\rb^{3d}
   \int d\eta du dv \,e^{i\eta (k+u)+2iv\e^{-1}r}
\\
&&\left[\int dy dn dm \,e^{i(u-v)y-i(u+v)(n+m)}\oe_{\e^{-1}t}[a^*(m) a(y)]
  {h_\e(n+m,m)} \right.
\\
&&-\left.\int dx dn dm \,e^{-i(u+v)x+i(u-v)(n+m)}\oe_{\e^{-1}t}[a^*(x) a(m)]
  \ol{h_\e(n+m,m)}
\right]
\\
&=&2^d\0 i \e \lb \0 1 {2\pi}\rb^{2d}
   \int dy dn dm  \, \oe_{\e^{-1}t}[a^*(m) a(y)] {h_\e(n+m,m)}
\\
&&\hspace{1em}\times \int du dv \,\delta(k+u) e^{-iu(n+m-y)} e^{2iv(\e^{-1}r-\0 {n+m+y} 2)}
\\
&-&2^d\0 i \e \lb \0 1 {2\pi}\rb^{2d}
   \int dx dn dm  \, \oe_{\e^{-1}t}[a^*(x) a(m)] \ol{h_\e(n+m,m)}
\\
&&\hspace{1em}
  \times\int du dv \,\delta(k+u) e^{iu(n+m-x)} e^{2iv(\e^{-1}r-\0 {n+m+x} 2)}
\\
&=&\0 {-i} \e \lb \0 1 {2\pi}\rb^{d}
   \int dy dn dm  \, \oe_{\e^{-1}t}[a^*(m) a(y)] {h_\e(n+m,m)}
\\
&&\hspace{5em}
  \times e^{ik(n+m-y)} \delta(\e^{-1}r-\0 {n+m+y} 2)
\\
&+&\0 i \e \lb \0 1 {2\pi}\rb^{d}
   \int dx dn dm  \, \oe_{\e^{-1}t}[a^*(x) a(m)] \ol{h_\e(n+m,m)}
\\
&&\hspace{5em}
  \times e^{-ik(n+m-x)} \delta(\e^{-1}r-\0 {n+m+x} 2)
\\
&=&\0 i \e \lb \0 1 {2\pi}\rb^{d}
  \int dn d\bar{y} d\eta  \, \oe_{\e^{-1}t}[(a^*(\bar{y}+\0 \eta 2) a(\bar{y}-\0 \eta 2)]
   {h_\e(n+\bar{y}+\0 \eta 2,\bar{y}+\0 \eta 2)}
\\
&&\hspace{5em}
  \times e^{ik(n+\eta )}\delta(\e^{-1}r-\0 n 2-\bar{y})
\\
&-&\0 i \e \lb \0 1 {2\pi}\rb^{d}
   \int dn d\bar{x} d\eta  \, \oe_{\e^{-1}t}[(a^*(\bar{x}+\0 \eta 2)
   a(\bar{x}-\0 \eta 2)]
   \ol{h_\e(n+\bar{x}-\0 \eta 2,\bar{x}-\0 \eta 2)}
\\
&&\hspace{5em}
  \times e^{-ik(n- \eta )}\delta(\e^{-1}r-\0 n 2-\bar{x})
\\
&=&\0 i \e \lb \0 1 {2\pi}\rb^{d}
   \int dn d\eta \, \left[ e^{ik(n+\eta)}
   \oe_{\e^{-1}t}[a^*(\e^{-1}r-\0 n 2+ \0 \eta 2)a(\e^{-1}r-\0 n 2- \0 \eta 2)]
   \right.
\\
&&\hspace{5em}\times
  {h_\e(\e^{-1}r+\0 {n+\eta} 2,\e^{-1}r- \0 {n-\eta} 2)}
\\
&-&\0 i \e \lb \0 1 {2\pi}\rb^{d}
   \int dn d\eta \,\left. e^{-ik(n-\eta)}
   \oe_{\e^{-1}t}[a^*(\e^{-1}r-\0 n 2+ \0 \eta 2) a(\e^{-1}r-\0 n 2- \0 \eta 2)]
   \right.
\\
&&\hspace{5em} \left.
  \times \ol{h_\e(\e^{-1}r+\0 {n-\eta} 2,\e^{-1}r- \0 {n+\eta} 2)}
   \right].
\eeann

Now, inserting the definitions of $f^\e(r,k;t)$ and $h(r,n)$ into the
last expression and expanding both at $r$, we find that

\beann
&&\lim_{\e\da0}\p_t f^\e(r,k;t)\\
&&=\lb \0 1 {2\pi}\rb^{d} \lim_{\e\da0}\0 i \e \int dn d\eta dv \,
   e^{ik\eta -i\eta v} f^\e(r-\e \0 n 2,v;t)
   \Big[e^{ikn} \,{h(r+\e\0 \eta 2,-n)}-e^{-ikn}\,
   {h(r-\e\0 \eta 2,n)}
   \Big]\\
&&=\lb \0 1 {2\pi}\rb^{d}\lim_{\e\da0}\0 i \e \int  d\eta dv \,
   e^{ik\eta -i\eta v} f^\e(r,v;t) \int dn\,
   \Big(e^{ikn} \,{h(r,-n)}-e^{-ikn}\,
   {h(r,n)}\Big)\\
&&+\lb \0 1 {2\pi}\rb^{d}
  \lim_{\e\da0}\int d\eta dv \,e^{ik\eta-i\eta v} \p_r f^\e(r,v;t)
  \int dn\, i n \,e^{-ikn} h(r,n)\\
&&+\lb \0 1 {2\pi}\rb^{d}
  \lim_{\e\da0}\int d\eta dv \,i\eta e^{ik\eta-i\eta v} f^\e(r,v;t)\,\p_r
  \int dn\,  e^{-ikn} h(r,n)\\
&&=-\p_k E(r,k)\p_r f(r,k;t)+\p_r E(r,k)\p_k f(r,k;t).
   \qquad\Box
\eeann

\mysection{Remarks}

\ben

\item Proposition (\ref{prop1}) can be extended to time dependent quasi-free
time evolutions, $U(t,0)=T\Big[e^{-i\int_0^t dt'\, h_\e (t')}\Big]$, where
$h_\e (t)$
is a family of self-adjoint operators such that the integral gives a well-defined
self-adjoint operator. $T$ is the time ordering operation. If
$h_\e (x,y;t)=h(\e (x+y)/2,y-x;\e t)$ then the function
$E(r,k)$ in equation (\ref{euler}) is simply replaced by an analogous
function
$E(r,k;t)$.

\item The general solution of (\ref{euler}) is given by
\beq \label{class equ}
     f(r,k;t)=f(r(-t),k(-t);0).
\eeq
$r(t), k(t)$ are determined by the classical Hamilton's equations of motion:
\bea \label{hamilton1}
     \p_t r&=&\p_k E(r,k)\\
     \label{hamilton2}
     \p_t k&=&-\p_r E(r,k)
\eea
with Hamilton function $E(r,k)$ and $r(0)=r, k(0)=k$.

\item Analyzing a sum of (properly scaled) quasi-free Hamilton operators
is reduced to the sum of their respective energy spectra; applicable  for
example to the case
$h_{\e}=\0 1 {2m} \Big(\vec{p} + \0 e c \vec{A}(\e x)\Big)^2 + e V(\e x)$.


\item
The analysis can easily be extended to generators, $H$, which are not
neccessarily self-adjoint, but where the difference between $H$ and
its adjoint $H^*$ is of order $\e$.
As an example consider $h_\e=-\0 1 2 \De_\g$, generating a random walk
with bias $\g=(\g_1,\ldots,\g_d)$ on the lattice $\bZ^d$ (with unit
vectors $e_i$ in the positive lattice directions):
\[\De_\g
f(x)=\sum_{i=1}^d \Big[(1+\g_i/2) f(x + e_i)+(1-\g_i/2) f(x -
e_i)- 2f(x)\Big].
\]
If $\g_i=\g(\e)$ be such that $\g_i(\e)/\e\to a_i$  as
$\e\da0$. Then, $f(r,k;t)$ satisfies
\beq \p_t f(r,k;t)+\sin{k}\cdot[(\nabla+a)f(r,k;t)]=0.
\eeq
The velocity term comes from the real part of the dispersion relation
$E(r,k)=1-\cos{k}$ whereas the drift $a$ stems from the imaginary part
of $E$.

\item
Notice that the density of particles, $\r(r,t)$, is
given by \[\r(r,t)=\int dk\, f(r,k;t).
\]
An example where one can say more about $\r(r,t)$ is obtained from taking
as the initial distribution the function $f(r,k)=\0 1 {2\pi} \r(r)$ with
$k\in
[-\pi,\pi]$.  Take $\bZ^1$ for the lattice case,
$h=-\0 1 2 \De$ the lattice Laplacian
as one particle Hamiltonian and
the following initial ``diagonal'' states $\oe$:
\[\oe[a^*(x)a(y)]=\delta(x,y)\cdot \r(\e x).
\]
Then, $\r(r,t)$ satisfies the modified Bessel equation
\beq \label{bes}
\Big[ t^2 \p_t^2 + t \p_t -t^2 \De\Big] \r(r,t)=0.
\eeq
In other words, $\tilde{\r}(p,t)$ is the (normal) Bessel function, which for
$p\not=0$ goes oscillating to 0 as $|t|$ tends to infinity.

\een

\mysection{Quantum Hard Core Bosons}

In this Section we consider a simple model of interacting Bosons on $\bZ^1$
(cf.\cite{AB}). The pair interaction is the so-called on-site hard core
interaction (i.e.~hard core radius 0), which
has the exclusion effect that at most
one Boson occupies a single lattice site.

Let $H_\La(\la)$ be the following Boson Hamilton operator on
an interval $\La\subset\bZ$
with point interaction $V(x,y)=\la \cdot \de(x,y)$ of strength $\la$
and empty boundary conditions,

\beq H_\La(\la)=-\s0 1 2 \sum_{x\in\La} b^*(x)(\De b)(x)+
                \la \sum_{x\in\La}b^*(x)^2 b(x)^2.
\eeq
Here, $b^*(x),b(y)$ denote the Bose field operators on $\bZ$. $H_\La(\la)$
acts on the usual Bosonic Fock space, ${\cal F}(\La)$. The limit of
$H_\La(\la)$ as $\la$ goes to infinity displays an interesting
Hamilton operator. This limit is well-defined on states
\beq |\psi_X\rangle=\prod_{x\in X} b^*(x)|0\rangle,\qquad X\subset\La,
\eeq
which define the Fock space, ${\cal F}^{hc}(\La)$, of hard core Bosons.
$|0\rangle$ is the vacuum state. Let $P: {\cal F}(\La)\to
{\cal F}^{hc}(\La)$ be the orthogonal projection onto the hard core Boson states.
We introduce the following annihilation (creation) operators
\beq a^{(*)}(x)=Pb^{(*)}(x)P.
\eeq
In particular, they have the following properties:

\ben

\item The state space ${\cal F}^{hc}(\La)$ is invariant under $b(x)$ and
therefore $a(x)=b(x)P$.

\item $a^*(x),a(y)$ satisfy mixed (Anti)Commutation relations:
\bea {[}a^*(x),a(y){]}_{-}&=&0, \quad \mbox{for } x\not= y,\\
     {[}a^*(x),a(x){]}_{+}&=&1,
\eea
where $[,]_{-}, [,]_{+}$ denote the Anti-, commutator, respectively.

\item $\lim_{\la\to\infty} H_\La(\la)=-\s0 1 2 \sum_{x\in\La}
      a^*(x)(\De a)(x)$ on ${\cal F}^{hc}(\La)$.

\een

The commutation relations are familiar from spin systems. Although the previous
formulation works in any dimension the statistics of the $a^\#(x)$'s can
only be efficiently
``repaired'' in one dimension. By applying the Klein-Jordan-Wigner
transformation
(cf.\cite{LSM}) we map the hard core Bose field operators onto Fermi field operators.

\begin{definition} Let
\bea c(x)&=&e^{-i\pi\sum_{j\le x-1} a^*(j)a(j)} a(x),
\\
     c^*(x)&=&a^*(x)e^{i\pi\sum_{j\le x-1} a^*(j)a(j)},
\eea
where the sum goes over all $j\in\La$ left to $x$. Then, $c^*(x),c(y)$ satisfy the
usual Anticommutation relation.
\end{definition}
Observe that under the above transformation, formally,
\beq\label{XY} H_\bZ=-\s0 1 2 \sum_{x\in\bZ} a^*(x)(\De a)(x)
                    =-\s0 1 2\sum_{x\in\bZ} c^*(x)(\De c)(x)
\eeq
is the free Fermion Hamilton operator. We know that
\beq n(r,k)=\sum_{\eta\in\bZ} e^{2 i\eta k} c^*(r-\eta) c(r+\eta)
\eeq
are the locally conserved quantities for the free force evolution. Rewriting
them in terms of hard core Bosons we get
\beq n(r,k)=\sum_{\eta\in\bZ} e^{2 i\eta k} e^{N(r-\eta,r+\eta)}
     a^*(r-\eta) a(r+\eta),
\eeq
where $N(x,y)=\sum_{x<j\le y} a^*(j)a(j)$. Although the time evolution of the $a^\#(x)$'s
is quite complicated (and unknown) we are only interested in a special combination
appearing in $n(r,k)$. The problem is then simply reduced to free Fermions on the lattice
(but only as far as this local quantity, $n(r,k)$, is
concerned).  It is then a simple matter to apply Proposition 1 :

\begin{prop} Suppose that $f(r,k) = \lim_{\e\da0} \oe[n(\e^{-1}r,k)]$
exists (for example if
the functions $\psi^{\e}_{\la}$ in the
decomposition of $\oe$ (see \ref{decomp}) are $\e$-oscillatory (see
\cite{GMMP}) and bounded).
Then, under the dynamics generated by the Hamilton operator (\ref{XY}),
$f(r,k;t)$ exists
and satisfies the free force Euler equation
\beq \p_t f(r,k;t)+\sin{k}\cdot\p_r f(r,k;t)=0
\eeq
\end{prop}

\noindent{\bf Remarks:} \ben

\item Since we are working in an infinite discrete configuration space we can map any
configuration of Bosonic hard rods with positive (integral) length
bijectively onto a
configuration of Bosons with core length 0, i.e.~the above case. Therefore
the last
proposition extends to hard core interactions of positive length.

\item In continuous configuration space things are much more delicate.
Whereas the same
analysis is literally applicable to Bosons on $\bR^1$ with core radius 0,
the hard rod model with Neumann boundary conditions (for example) seems to be very
different and rather complicated.
\een

\mysection{Fluctuations of the Density}

The Euler equation corresponds mathematically to a law of large numbers.
The simplest equation for the density profile is (\ref{bes}).
Now we study (non-equilibrium) fluctuations of the particle density for
the
quasi-free systems of Section 3. The field of fluctuations,
$\xi^\e(\psi,A,t)$, of a local observable, $A$, is defined by

\beq \label{fluc}
\xi^\e(\psi,A,t)=\e^{d/2} \int dr\,
     \psi(\e r)\left[ \tau_{r,\e^{-1}t}(A)-\oe(\tau_{r,\e^{-1}t}(A)\right].
\eeq
Here, $\psi$ is any test function and $\oe$ a family of states as in Section 2.
Further, we denote by $\tau_{r,t}(A)$ the
space and time shift of the observable, $A$, by $r$ and $t$, respectively.
We are interested in the density, $\tau_{r,t}(A)=\a_t(a^*(r)a(r))$, in
particular in realizing the limit

\beq \lim_{\e\da0}\xi^\e(\psi,A,t)=
     \int dr\,\psi(r)\xi(r,t).
\eeq
We should therefore study the characteristic function, $\oe\lb
e^{i\la\xi^\e (\psi,A,t)}\rb$, and try to reconstruct the limiting
dynamics of the fluctuation field. Since we restrict our attention to
quasi-free states (cf.\cite{BR}) and we always assume nice clustering
properties it is fair to expect Gaussian behavior.  The precise meaning
of this in the quantum case (the so called quantum central limit theorem)
can be found in  \cite{GVV1} and \cite{GVV2}. We are satisfied therefore
with studying the covariance $\lim_{\e\da0} \oe[\xi^\e(\psi,t)\xi^\e(\phi,s)]$
with  $\xi^\e(\phi,s)$ given by (\ref{fluc}) for $A=a^*(0)a(0)$ and for a
class of test functions $\psi,\phi \in \cal D$.

Clearly, the dynamics of the fluctuations can only be derived when
the law of large number is already established.  Therefore, while
the next proposition does not follow directly from Proposition 1,
it is essential to make sense of fluctuations that one first
understands the averaged behavior.  We put ourselves therefore in
the same context as for Proposition 1 and we have that $f(r,k;t)
=\lim_{\e\downarrow 0} f^\e(r,k;t)$. On the level of the
one-particle dynamics we recall that we denote by $r^\e(t)$ the
rescaled position operator, see (2.3), and we assume that it
converges to an operator $r(t)$ as $\e\da0$, see \cite{MMP},
\cite{K}, \cite{S3}.  This convergence takes place on a suitable
domain $\cal D$ dense in $L^2(\bR^d,dx)$. Let us write
\beq\lb e^{ir(t)}\psi\rb(x,k)=\lb \0 1 {2\pi}\rb^{d/2}
                            \int dq\,\tilde{\psi}(q) e^{iqr(x,k;t)},
\eeq
where $r(x,k;t)$ is the solution of (\ref{hamilton1}).
(If the motion is force free then $\lb e^{ir(t)}\psi\rb(x,k) = \psi(x+
\partial_k E(k) t)$.)

\begin{prop} Under the above conditions on the initial states and on
the Hamilton operator,
\beq \lim_{\e\da0} \oe[\xi^\e(\psi,t)\xi^\e(\phi,s)]=
     \int dx dk\, \lb e^{ir(t-s)}\psi\rb(x,k) \phi(x) f(x,k;s)(1\pm f(x,k;s)),
\eeq
where $+/-$ concerns Fermi/Bose statistics, respectively.
\end{prop}

\noindent{\bf Proof:}

\beann \lefteqn{\oe[\xi^\e(\psi,t)\xi^\e(\phi,s)]
}\\
&=& \e^d \int dr dr' \, \psi(\e r)\phi(\e r') \Big\{
   \oe[a^*(r,\e^{-1}t) a(r,\e^{-1}t) a^*(r',\e^{-1}s) a(r',\e^{-1}s)]
\\
&-&\oe[a^*(r,\e^{-1}t) a(r,\e^{-1}t)]\,\oe[a^*(r',\e^{-1}s) a(r',\e^{-1}s)]
   \Big\}
\\
&=& \e^d \int dr dr' \, \psi(\e r)\phi(\e r')
   \oe[a^*(r,\e^{-1}t) a(r',\e^{-1}s)] \, \oe[a(r,\e^{-1}t)a^*(r',\e^{-1}s)].
\eeann

By changing the order of $a$ and $a^*$ in the last expectation we get two parts. The one
including the (anti) commutator is equal to

\beann \lefteqn{ \e^d \int dr dr' \, \psi(\e r)\phi(\e r')
   \oe[a^*(r,\e^{-1}t) a(r',\e^{-1}s)]\,\oe\lb[a(r,\e^{-1}t),a^*(r',\e^{-1}s)]_\pm\rb
}\\
&=&\lb \0 1 {2\pi}\rb^{d/2}
    \e^d \int dr' dq \, \tilde{\psi}(q)\phi(\e r')
   \int dz dr \, e^{iq\e r}\, e^{-ih_\e(t-s)\e^{-1}}(r,z) \,
   e^{ih_\e(t-s)\e^{-1}}(r',r)
\\
&&\hspace{6em} \times\oe[a^*(z,\e^{-1}s)a(r',\e^{-1}s)]
\\
&=&\lb \0 1 {2\pi}\rb^{d/2}
    \e^d \int dr' dq \, \tilde{\psi}(q)\phi(\e r')
   \int dz \,e^{iq r^\e(t-s)}(r',z) \,\oe_{\e^{-1}s}[a^*(z)a(r')]
\\
&=&\lb \0 1 {2\pi}\rb^{d/2}
    \e^d \int dx d\eta dq \, \tilde{\psi}(q)\phi(\e (x-\0 \eta 2))
   e^{iq r^\e(t-s)}(x-\0 \eta 2,x+\0 \eta 2) \,
   \oe_{\e^{-1}s}[a^*(x+\0 \eta 2)a(x-\0 \eta 2)]
\\
&=&\lb \0 1 {2\pi}\rb^{d/2}
   \int dx d\eta dk dq \, \tilde{\psi}(q) \phi (x-\e \0 \eta 2)
   e^{iq r^\e(t-s)}(\e^{-1}x-\0 \eta 2,\e^{-1}x+\0 \eta 2) \,e^{-i\eta k}
   f^\e(x,k;s)
\eeann

Now
remember our scaling of the matrix elements of $h_\e$ and take the limit
$\e\da0$  of the last expression:

\beann
\lefteqn{
\lb \0 1 {2\pi}\rb^{d/2}
   \int dx dk dq \, \tilde{\psi}(q) \phi (x)
   \lb\int d\eta \, e^{iq r(t-s)}(x,\eta) \,e^{-i\eta k}\rb
   f(x,k;s) }\\
&=&\int dx dk \, \lb e^{ir(t-s)}\psi\rb (x,k) \phi (x) f(x,k;s).
\eeann

For the second part (forgetting the $\lb {2\pi}\rb^{d/2}$) we need

\beann\lefteqn{
    \lim_{\e\da0}\e^d\int dq dr'\,\tilde{\psi}(q)\phi(\e r')\int dr \,e^{iq\e r}
    \oe_{\e^{-1}s}[a^*(r,\e^{-1}(t-s)) a(r')] \,
    \oe_{\e^{-1}s}[a^*(r') a(r,\e^{-1}(t-s))]
}\\
&=&\lim_{\e\da0}\e^d\int dq dr'\,\tilde{\psi}(q)\phi(\e r')\int dr dz dz'\,e^{iq\e r}
   e^{-ih\e^{-1}(t-s)}(r,z)\,\oe_{\e^{-1}s}[a^*(z) a(r')]
\\
&&\hspace{3em}\times e^{ih_\e (t-s)\e^{-1}}(z',r)\, \oe_{\e^{-1}s}[a^*(r') a(z')]
\\
&=&\lim_{\e\da0}\e^d\int dq dr'\,\tilde{\psi}(q)\phi(\e r')
   \int dz dz'\,e^{iq\e r(\e^{-1}(t-s))}(z',z) \,
   \oe_{\e^{-1}s}[a^*(z) a(r')] \,
   \oe_{\e^{-1}s}[a^*(r') a(z')]
\\
&=&\lim_{\e\da0}\e^d\int dq dx d\eta\,\tilde{\psi}(q)\phi(\e (x-\0 \eta 2))
   \int dz'\,e^{iqr^\e(t-s)}(z',x+\0 \eta 2) \,
   \oe_{\e^{-1}s}[a^*(x+\0 \eta 2) a(x-\0 \eta 2)]
\\
&&\hspace{3em}\times\oe_{\e^{-1}s}[a^*(x-\0 \eta 2) a(z')]
\\
&=&\lim_{\e\da0}\e^d\int dq dx d\eta\,\tilde{\psi}(q)\phi(\e x)
   \int  dz' dk \,e^{iqr^\e(t-s)}(z',x+\0 \eta 2)\,
   e^{-i\eta k}f^\e(\e x,k;s)
\\
&&\hspace{3em}\times\oe_{\e^{-1}s}[a^*(x-\0 \eta 2) a(z')]
\\
&=&\lim_{\e\da0}\e^d\int dq dy d\eta d\bar{\eta} dk d\bar{k}\,
   \tilde{\psi}(q)\phi(\e (y+\0 {\bar{\eta}} 2))
   \,e^{iqr^\e(t-s)}(y-\0 {\bar{\eta}} 2,y+\0 {\bar{\eta}} 2+\eta)
\\
&&\hspace{3em}\times e^{-i\eta k-i\bar{\eta} \bar{k}} f^\e(\e(y+\0 {\bar{\eta}+\eta} 2),k;s)
   f^\e(\e y,\bar{k};s)
\\
&=&\lim_{\e\da0}\int dq dy d\eta d\bar{\eta} dk d\bar{k}\,
   \tilde{\psi}(q)\phi(y+\e\0 {\bar{\eta}} 2)
   \,e^{iq r^\e(t-s)}
   (\e^{-1}(y+\e\0 {\eta} 2)-\0 {\eta+\bar{\eta}} 2,
    \e^{-1}(y+\e\0 {\eta} 2)+\0 {\eta+\bar{\eta}} 2 )
\\
&&\hspace{3em}\times e^{-i\eta k-i\bar{\eta} \bar{k}} f^\e(y+\e\0 {\bar{\eta}+\eta} 2),k;s)
   f^\e(y,\bar{k};s)
\\
&=&\lim_{\e\da0}\int dq dy d\eta d\bar{\eta} dk d\bar{k}\,
   \tilde{\psi}(q)\phi(y+\e\0 {\bar{\eta}} 2)
   \,e^{iq r(t-s)}
   (y+\e\0 {\eta} 2,\0 {\eta+\bar{\eta}} 2 )
\\
&&\hspace{3em}\times e^{-i\eta k-i\bar{\eta} \bar{k}} f^\e(y+\e\0 {\bar{\eta}+\eta} 2),k;s)
   f^\e(y,\bar{k};s)
\\
&=&\lim_{\e\da0}\int dq dy d\nu d\bar{\nu} dp d\bar{p}\,
   \tilde{\psi}(q)\phi(y)
   \,e^{iq r(t-s)}(y,\nu)
   e^{-i\nu p-i\bar{\nu} \bar{p}} f^\e(y,p+\bar{p};s)
   f^\e(y,p-\bar{p};s)
\\
&=&\lb {2\pi}\rb^{d/2}\int dy d\bar{\nu} dp d\bar{p}\,
   \lb e^{ir(t-s)}\psi\rb (y,p)\phi(y)
   e^{-i\bar{\nu} \bar{p}} f(y,p+\bar{p};s)
   f(y,p-\bar{p};s)
\\
&=&\lb {2\pi}\rb^{d/2}\int dy dp \,\lb e^{ir(t-s)}\psi\rb(y,p)\phi(y)
   f^2(y,p;s).\qquad\Box
\eeann

\mysection{Particles in a Periodic Potential}

So far we have not dealt with the important case of particles moving in a
periodic potential. More precisely, where the system is quasi-free with one
particle Hamiltonian of the form $h=-\De+U$ on $\bR^d$ with periodic
potential $U$. Recently, Spohn (\cite{S3}) studied their long time behavior.
To obtain the corresponding Euler equations in this case is
clearly of interest but some problems immediately arise.

The first one is related to the identification and the proper
interpretation of the corresponding Wigner distribution function.  In
other words, by what to replace the equations (\ref{wigner})-(\ref{bbb})?
We do not know a unique good answer to that question (but see the
discussion in \cite{GMMP}, \cite{MMP}, \cite{K} and \cite{AK}).  An obvious generalization
of (2.4) goes as follows.

Let $\psi_{nk}$ be the eigenfunctions of the Hamiltonian
$H=-\De+U$, with periodic potential $U$, having
eigenvalues $E_n(k)$. Introduce
the generalized Fourier transformation (cf.\cite{RS})
\beq \label{gen fourier}
     \tilde{\chi}(k,n)=\int_{\bR^d} dx\, {\psi}_{nk}(x) \chi(x).
\eeq

We consider now an ensemble (we denote creation and annihilation operators by $a^*(x),
a(y)$) of particles subject to a periodic potential $U$ and a
family of states $\oe$ for which the following limit exists
($\La^*$ is the first Brillouin zone)
\beq f_n(r,k)=\lim_{\e\da0} \e \int_{\bR^d} dv\,e^{-irv}\oe[\tilde{a}^*(k+\e \0 v 2,n)
              \tilde{a}(k-\e \0 v 2,n)],\qquad r\in\bR^d,k\in\La^*
\eeq
and call it the $n$-th band distribution function of the ensemble.
$\tilde{a}(k,n)$
is to be understood as (\ref{gen fourier}), but now as operators. We
recover (\ref{wigner}) at time $t=0$ if in (7.1)-(7.2) we use
$\psi_{nk}(x)=\0 1 {2\pi} e^{-ikx}$.  Comparing with (\ref{aaa})-(\ref{bbb}),
we have
\beq \int dk f_n(r,k) = \lim_{\e\da0} \oe (a^*(\e^{-1}r,n)a(\e^{-1}r,n))
\eeq
and
\beq \int dr f_n(r,k) =  \lim_{\e\da0} \e \oe (\tilde{a}^*(k,n)
\tilde{a}(k,n))
\eeq

\begin{prop} Let $\a_t(\cdot)$ be the time evolution of the ensemble generated by the
one particle Hamilton operator
$H=-\De+U$ on $\bR^d$: $H\psi_{nk}=E_n(k)\psi_{nk}$.
Then,
\beq f_n(r,k;t)=\lim_{\e\da0} \e \int_{\bR^d}
dv\,e^{-irv}(\oe\circ\a_{\e^{-1}t})
                [\tilde{a}^*(k+\e \0 v 2,n) \tilde{a}(k-\e \0 v 2,n)]\,
                \qquad r\in\bR^d,k\in\La^*
\eeq
exists and
\beq \p_t f_n(r,k;t)+\partial_k E_n(k)\cdot\p_r f_n(r,k;t)=0.
\eeq
\end{prop}

\noindent{\bf Proof:}
This is an easy task since $\a_t \tilde{a}(k,n)=e^{-itE_n(k)}
\tilde{a}(k,n)$ and $f_n(r,k;t) = f_n(r - \partial_k E_n(k) t,k).
\Box
$

\bigskip

\noindent
{\bf Acknowledgment:}
We have very much benefited from discussions with B.~Baumgartner, M.~Fannes,
M.~Moser, O.~Penrose, especially with H.~Spohn, Yu.~Suhov and
A.~Verbeure. We are also indepted to the referee who draw our
attention to \cite{GMMP}. WS was supported
by the project P10517-NAW of the \"Osterreichische Forschungsfond, FWF. He also
wants to thank the Institute of Theoretical Physics in Leuven for their very warm
hospitality and financial support.


\begin{thebibliography}{Dillo 99}

\bibitem[AB]{AB} N.~Angelescu and M.~Bundaru: {\em A Remark on the Condensation
   in the Hard-Core Lattice Gas}, J.~Stat.~Phys., {\bf 69}, 897 (1992)

\bibitem[AK]{AK} Ash and Knauf, mp-arc 97-545

\bibitem[AVZ]{AVZ}  N.~Angelescu, A.~Verbeure and V.A.~Zagrebnov:
   J.~Phys. A, {\bf 25}, 3473 (1992)

\bibitem[B]{Bog} N.N.~Bogoliubov: J.~Phys. (USSR), {\bf 11},
   23 (1947)

\bibitem[BDS]{BDS} C.~Boldrighini, R.L.~Dobrushin and Yu.M,~Suhov: {\em
   One-dimensional hard rod caricature of hydrodynamics}, J.~Stat.~Phys.,
   {\bf 31}, 577 (1983)

\bibitem[BS]{BS} C.~Boldrighini and Yu.M,~Suhov: {\em
   One-dimensional hard rod caricature of hydrodynamics}: {\em ``Navier-Stokes
   Correction'' for Local Equilibrium States}, Comm. Math. Phys.,
   {\bf 189}, 577 (1997)


\bibitem[BR]{BR} O.~Bratteli and D.W.~Robinson: {\em Operator Algebras and
   Quantum Statistical Mechanics II. Equilibrium States Models in Quantum
   Statistical Mechanics}, Springer (1981)

\bibitem[D]{Dob} R.L.~Dobrushin: {\em Caricatures of hydrodynamics}. In :
   IXth International Congress on Mathematical Physics, eds. B. Simon, A.
   Truman, I.M. Davies, pp. 117-132. Adam Hilger, Bristol (1989)

\bibitem[GMMP]{GMMP} P.~Gerard, P.A.~Markowich, N.J.~Mauser and
   F.~Poupaud: {\em Homogenization Limits and Wigner Transforms}, Comm.
   Pure and Appl. Math, Vol L, 323 (1997)


\bibitem[GVV1]{GVV1} D.~Goderis, A.~Verbeure and P.~Vets:
  {\em Non-commutative central limits}, Prob. Th. Rel. Fields {\bf 82}, 527
  (1989)


\bibitem[GVV2]{GVV2} D.~Goderis, A.~Verbeure and P.~Vets:
  {\em Central Limit Theorem for mixing Quantum Systems and the CCR-algebra
  of Fluctuations}, Comm. Math. Phys. {\bf 122}, 249 (1989)

\bibitem[K]{K} A. Knauf: Comm. Math. Phys. {\bf 109}, 1 (1987)

\bibitem[LSM]{LSM} E.H.~Lieb, T.~Schultz and D.~Mattis: {\em Two
  soluble Models of an Antiferromagnetic Chain}, Ann.~Phys., {\bf 16},
  407 (1961)

\bibitem[LP]{LP} E.M.~Lifshitz and L.P.~Pitaevskii : {\em L.D. Landau
  and E.M. Lifshitz Course on Theoretical Physics}, Volume 9, Statistical
  Physics, Part 2, Pergamon Press Ltd (1980).

\bibitem[MMP]{MMP} P.A.~Markowich, N.J.~Mauser and F.~Poupaud: {\em A
  Wigner-function approach to
 (semi)
   classical limits: Electrons in a periodic
  potential}, J. Math. Phys. {\bf 35}, 1066--1094 (1994)

\bibitem[RS]{RS} M.~Reed and B.~Simon: {\em Methods of Modern Mathematical
  Physics, IV: Analysis of Operators}, (1978)

\bibitem[S1]{S1} H.~Spohn: {\em Large Scale Dynamics of Interacting
  Particles}, Springer, (1991)

\bibitem[S2]{S2} H.~Spohn: {\em Quantum Kinetic equations}, {\bf 1},
  "On three levels", ed. by M.~Fannes et al, (1994)

\bibitem[S3]{S3} H.~Spohn: {\em Long Time Asymptotics for Quantum Particles
  in a Periodic Potential}, Phys.~Rev.~Lett., vol 77, nr 7, 1198, (1996)

\bibitem[W]{Wig} E.~Wigner: {\em On the Quantum Correction
  for Thermodynamic Equilibrium}, Phys.~Rev., {\bf 40}, 749 (1932)


\end{thebibliography}
\end{document}